\documentstyle[preprint,aps]{revtex}
\tightenlines

\newcommand{\ket}[1]{|{#1}\rangle}
\newcommand{\th}{\theta}

\begin{document}

\title{Entanglement Swapping using Continuous Variables}
\author{R.E.S.Polkinghorne and T.C.Ralph}
\address{Department of Physics, Faculty of
Science, \\ The Australian National University, \\ ACT 0200 Australia \\ 
 E-mail: Timothy.Ralph@anu.edu.au}
\maketitle

\begin{center}
\scriptsize (submitted to PRL 22nd October 1998)
\end{center}

\begin{abstract}

We investigate the efficacy with which entanglement can be 
teleported using a continuous measurement scheme. We show that by 
using the correct gain for the classical channel the degree of violation 
of locality that can be demonstrated (using a CH type inequality) 
is {\it not} a function of the 
level of entanglement squeezing used in the teleportation. This is 
possible because a gain condition can always be choosen such that passage through 
the teleporter is equivalent to pure attenuation of the input field.

\end{abstract}

\vspace{10 mm}

It is remarkable that non-local entanglement can be 
established between particles that have never interacted directly. 
Here ``non-local'' refers to the inability of local hidden variable 
theories to predict the observed correlations.
This ``entanglement swapping'' \cite{zuk93}, \cite{ben93} may be useful in 
establishing non-local 
correlations over very large distances and other applications 
\cite{bos98}. Recently 
Pan et al \cite{zei98} have demonstrated entanglement swapping 
of the polarization entanglement created by type II 
parametric down conversion experimentally. 
In all discussions and experiments to date discrete 
measurements and manipulations are made in order to transfer the 
non-local correlations. For example in the optical experiments, photon 
coincidences operate photo-current gates.
However entanglement swapping is really a special case of 
teleportation \cite{ben93} and in work by Vaidman \cite{vai94} 
and Braunstein and Kimble \cite{bra98}, 
schemes 
for the teleportation of continuous quantum 
variables have been proposed. In these schemes continuous measurements 
and manipulations are used. A preliminary experimental demonstration 
of continuous variable teleportation of a coherent state has recently 
been presented by Furusawa et al \cite{fur98}.
An important question to ask is; can 
non-local entanglement be swapped or teleported using a continuous 
measurement scheme? 

In this paper we show explicitly that this can be achieved. 
This effect represents 
a completely new way of transferring non-local information. Of 
particular practical significance is that the conditions for achieving 
non-local effects are not stringent.

The optical arrangement we will investigate is shown in Fig.~1. 
It combines the basic arrangement of entanglement swapping 
\cite{zei98} with a 2-mode generalization of the continuous variable 
teleportation scheme \cite{bra98}. 
We consider a non-collinear type II optical parametric oscillator operating 
at low pump efficiency (OPO1) as our source of entangled 
photons \cite{kwi95}. 
In the Heisenberg picture the two outputs, $A$ and $B$, can be 
decomposed into their horizontal ($h$) and vertical ($v$) linear 
polarization components by
\begin{eqnarray}
  A = 
   A_{(h)}\hat h+ A_{(v)}\hat v \nonumber\\
  B =
   B_{(h)}\hat h+ B_{(v)}\hat v
   \label{OPb}
\end{eqnarray}
where $\hat h$ and $\hat v$ are orthogonal unit vectors,
\begin{eqnarray}
  A_{(h,v)} = 
   A_{0(h,v)} \cosh\chi_{1} + B_{0(v,h)}^{\dagger} 
   \sinh\chi_{1}, \nonumber\\
  B_{(h,v)} =
   B_{0(h,v)} \cosh\chi_{1} + A_{0(v,h)}^{\dagger} \sinh\chi_{1},
   \label{OPp}
\end{eqnarray}
$A_{0}$ and $B_{0}$ are the vacuum inputs to OPO1, and 
$\chi_{1}$ is its conversion efficiency. We have assumed the bandwidth of 
the OPO is broad compared to our detection bandwidth and that pump 
depletion can be ignored.  The output state of the combined system in 
the number state basis is 
given by 
\begin{eqnarray}
  \frac{1}{\sqrt{2} \cosh(\chi_{1})} \sum_{n=0}^{\infty} 
  (\tanh\chi_{1})^{n} \left( \ket{n_{h},n_{v}} + \ket{n_{v},n_{h}} 
  \right)
\end{eqnarray}
where
\begin{equation}
\ket{n_{i},n_{j}}\equiv \ket{n_{i}}_{A}\otimes 
\ket{n_{j}}_{B}
\end{equation}
and $n_{h}$ and $n_{v}$ are the photon number in the 
horizontal and vertical polarizations respectively.

This reduces to the number-polarization entangled state
\begin{eqnarray}
  \frac{\chi_{1}}{\sqrt{2}}\left( \ket{1_{h},1_{v}} + \ket{1_{v},1_{h}} 
  \right)+\ket{0}
\label{state}
\end{eqnarray}
for low pump efficiency (i.e $ \chi_{1} \ll 1 $ ).  The state given by 
Eq~\ref{state} violates the Clauser-Horne (CH) type inequality 
\cite{cla74} 
\begin{eqnarray}
  S = \frac{ R(\th_{A}, \th_{B}) - R(\th_{A}, \th_{B}^{\prime}) 
           + R(\th_{A}^{\prime}, \th_{B}) 
           + R(\th_{A}^{\prime}, \th_{B}^{\prime}) }
           { R(\th_{A}^{\prime}, -_{B}) + R(-_{A}, \th_{B}) }
    \leq 1
\label{CH}
\end{eqnarray}
where $R(\th_{A}, \th_{B})$ is the photon coincidence count rate 
between polarisation $\th_{A}$ of beam $A$ and $\th_{B}$ of $B$, and 
$R(\th_{A}, -_{B})$ is the equivalent rate counting both polarisations of 
beam $B$.  The maximum violation occurs for $\th_{A} = \pi/8$, 
$\th_{B} = - \pi/4$, $\th_{A}^{\prime} = 3 \pi/8$ and 
$\th_{B}^{\prime} = 0$, when $S \approx 1.21$. Strong violations of 
locality have been observed experimentally with such a state 
\cite{kwi95}.

Now we consider teleporting, or swapping the entanglement of, 
one of the beams ($B$) from our non-local source 
using a continuous variable method. We will then investigate the 
correlations between the teleported beam and beam $A$ and determine under what 
circumstances they still violate the CH inequality. The teleportation 
is achieved using a second type II OPO (OPO2). The output beams of 
OPO2, $C$ and $D$, are given by analogous expressions to those of 
OPO1 (Eq.\ref{OPb},\ref{OPp}). The conversion efficiency of OPO2 is 
$\chi_{2}$. Beam $B$ is split into its two 
polarizations components ($B_{h}$ and $B_{v}$) 
at a polarizing beam-splitter (see Fig.~1). Similarly beam $C$ is split 
into $C_{h}$ and $C_{v}$. The horizontally polarized component of 
OPO1 ($B_{h}$) is mixed with the horizontally 
polarized component from OPO2, $C_{h}$, on a 50:50 beamsplitter. 
The outputs of the 
beamsplitter are directed to two homodyne detection systems which 
measure the 
phase ($X^{-}$) and amplitude ( 
$X^{+}$) quadratures of the field. 
Similarly $B_{v}$ and $C_{v}$ are mixed and their quadrature 
amplitudes detected. 
The resulting photocurrents are proportional to
\begin{equation}
  X_{(h,v)}^{\pm} = \sqrt{1-\eta} X_{\delta (h,v)}^{\pm} 
   + \sqrt{\eta/2}(X_{B(h,v)}^{\pm}
     \pm X_{C0(h,v)}^{\pm} \cosh\chi_{2} 
     + X_{D0(v,h)}^{\pm}\sinh\chi_{2})
\end{equation}
where, for example $X_{B}^{-}=i(B-B^{\dagger})$ and 
$X_{B}^{+}=B+B^{\dagger}$. The operators $X_{\delta (h,v)}^{\pm}$ 
come from vacuum modes 
introduced by losses in the homodyne systems, which are assumed to 
have efficiencies $\eta$.
The photo-currents are then amplified and fed-forward 
to the interferometric modulation systems (IMS) depicted in Fig.~2 
which act on the individual polarization components of the second 
beam from OPO2, $D_{h}$ and $D_{v}$. The photocurrents from the 
detection of the horizontally polarized beams are used to modulate 
$D_{v}$ whilst the photo-currents from the detection of the vertically 
polarized beams are used to modulate $D_{h}$. The effect of the IMS's are to 
displace the amplitudes of the beams by coupling in power from local 
oscillator beams (LO). The coupling is achieved via electro-optic 
modulators (EOM) in the interferometer arms. Provided the phase shifts 
($\phi_{v,h}$) 
introduced by the EOM's are small, the output of the IMS's ($D_{h}'$ 
and $D_{v}'$)
are given by
\begin{equation}
D_{(h,v)}'=D_{(h,v)}+\bar E\;\phi _{v,h}
\end{equation}
where $\bar E$ is the coherent amplitude of the LO. In general 
we have
\begin{equation}
\phi_{v,h}(t)=\int\limits_{0 }^t  {k^+
\left( u \right)X_{v,h}^+\left( {t-u} \right)}du+
\int\limits_{0 }^t  {k^-\left( u \right)X_{v,h}^-
\left( {t-u} \right)}du
\end{equation}
where $k^{\pm}$ contains various constants of proportionality as 
well as the time response of the feedforward electronics. 
However, if we restrict our attention to RF frequencies (relative 
to the local oscillator) for which the frequency
response of the electronics is flat we can set
\begin{equation}
  k^{\pm}(u) = {{1}\over{\sqrt{2}\bar E}} \lambda^{\pm} \delta(u)
\end{equation}
where $\lambda^{\pm}$ is the feedforward gain, and so
\begin{equation}
D_{(h,v)}'=D_{(h,v)}+{{1}\over{\sqrt{2}}}\lambda^{+} X_{v,h}^{+}
+{{1}\over{\sqrt{2}}}\lambda^{-} X_{v,h}^{-}
\end{equation}
Finally the beams are recombined using a polarizing beamsplitter and a 
half-wave plate is used to rotate horizontal polarizations 
into vertical and vice versa. The output beam is
\begin{equation}
D'=D_{(v)}' \hat h+D_{(h)}' \hat v
\end{equation}
Setting ($\lambda=-\lambda^{+}=i \lambda_{-}$) and assuming unit 
detection efficiency 
($ \eta=1 $) we obtain
\begin{eqnarray}
  D' & = & \left( B_{(h)} 
       + (\sinh\chi_{2} - \lambda \cosh\chi_{2}) C_{0h}^{\dagger}
       + (\cosh\chi_{2} - \lambda \sinh\chi_{2}) D_{0v} \right) \hat h
       \nonumber \\ 
     &   & + \mbox{} \left( B_{(v)} 
       + (\sinh\chi_{2} - \lambda  \cosh\chi_{2}) C_{0v}^{\dagger}
       + (\cosh\chi_{2} - \lambda  \sinh\chi_{2}) D_{0h} \right) \hat v
       \label{out}
\end{eqnarray}
In the limit of strong squeezing ($ \chi_{2} \gg 1$ such that 
$\cosh\chi_{2}\approx \sinh\chi_{2}$) and unity gain ($\lambda=1$) 
beams $B$ and $D'$ become equivalent.
It is clear that in this limit the beams $A$ and 
$D'$ will violate the CH inequality for conditions under which $A$ 
and $B$ violated it, showing that the non-locality has been teleported. 
This is shown in Fig.~3 where Eq~\ref{CH} is evaluated as a function of 
polarizer angle with beams $A$ and 
$D'$ as inputs, unity gain and 99\% squeezing. 

Very high levels of squeezing are difficult to achieve so it 
is important to ascertain what levels of squeezing are required to 
achieve non-local teleportation. Indeed, if we remain at unity 
gain, the operating point discussed in Ref. \cite{bra98} and used by 
Furusawa et al \cite{fur98}, Fig.~3 also 
shows that non-locality is lost for squeezing less than 
about 80\%. Surprisingly, though, we are able to recover non-local 
behavior for low levels of squeezing if we reduce the gain in the feedforward 
loops. This represents a new and potentially useful operating point.

We can write an analytical relationship between the value of $S$ that 
could be obtained from photon correlation measurements of beams $A$ 
and $B$, $S_{A,B}$ and that which 
could be obtained for the same measurements of beams $A$ and $D'$, 
$S_{A,D'}$ in the limit that $ \chi_{1} \ll 1 $. We must calculate 
photon coincidence count rates between beams $A$ and $D'$ such as
\begin{equation}
R(\theta_{A},\theta_{D'})=\langle in|E^{\dagger}_{D'}(\theta_{D'})
E^{\dagger}_{A}(\theta_{A})E_{A}(\theta_{A})E_{D'}(\theta_{D'})|in 
\rangle
\end{equation}
where
\begin{eqnarray}
 E_{A}(\theta_{A}) & = & A_{h}  cos\theta_{A}+ A_{v} 
 sin\theta_{A}\nonumber\\
 E_{D'}(\theta_{D'}) & = & D'_{h}  cos\theta_{D'}+ D'_{v} sin\theta_{D'}
 \end{eqnarray}
and $|in\rangle$ is given by Eq.~\ref{state}. After some algebra one 
finds
\begin{equation}
R(\theta_{A},\theta_{D'})=\lambda^{2} \eta R(\theta_{A},\theta_{B})+(N^{2}
+\lambda^{2}(1-\eta))/2
\end{equation}
where
\begin{equation}
N=\sinh\chi_{2} -\lambda \sqrt{\eta} \cosh\chi_{2}
\end{equation}
Similarly we find
\begin{equation}
R(\theta_{A},-_{D'})=\lambda^{2} \eta R(\theta_{A},-_{B})+(N^{2}
+\lambda^{2}(1-\eta))
\end{equation}
and
\begin{equation}
R(-_{A},\theta_{D'})=\lambda^{2} \eta R(-_{A},\theta_{B})+(N^{2}
+\lambda^{2}(1-\eta))
\end{equation}
Putting these results together as per Eq.~\ref{CH} we obtain
\begin{equation}
S_{A,D'}={{{{N^{2}}\over{\lambda^{2}}}+\eta S_{A,B}+1-\eta}\over
{{{2 N^{2}}\over{\lambda^{2}}}+2-\eta}}
\label{SAD}
\end{equation}
Consider first unit detection efficiency ($\eta=1$). 
Eq.~\ref{SAD} shows that the non-local correlation is preserved by 
the teleportation {\it for any level of squeezing}
provided we set 
\begin{equation}
\lambda_{op}=\tanh\chi_{2}
\label{gain}
\end{equation}
This effect is shown in Fig.~4 where the maximum of $S_{A,D'}$ is plotted 
against the feedforward gain, $\lambda$ for various levels of 
squeezing. As squeezing is reduced equal violations of locality are still 
achieved for lower levels of gain. The range of feedforward gains for 
which non-local teleportation is achieved actually broadens to a 
maximum value as the squeezing is reduced before narrowing again. The 
mechanism for this surprizing result can be understood by examining the 
action of the teleporter on an arbitrary, single mode input field, 
$a_{in}$. Under ideal conditions the output field is given by
\begin{equation}
a_{out}=\lambda a_{in}+(\cosh\chi-\lambda\sinh\chi)B_{0}-
(\lambda\cosh\chi-\sinh\chi)A_{0}^\dagger
\end{equation}
Notice that photons are added to the output through the action of the 
creation operator, $A_{0}^\dagger$. These spurious photons are detrimental 
to the observation of non-local correlations. However no photons are added 
to the output if the gain condition $\lambda_{op}=\tanh\chi$ is choosen as 
the coefficient of $A_{0}^\dagger$ goes to zero. The output is then given 
by
\begin{equation}
a_{out}=\lambda_{op} a_{in}+\sqrt{1-\lambda_{op}^2}B_{0}
\label{att}
\end{equation}
Eq.\ref{att} is formally equivalent to pure attenuation by a factor 
$(1-\lambda_{op}^{2})$. Thus when the teleporter is operated with this 
gain the output beam $D'$ is simply an attenuated version of $B$.
Because $S$ is a normalized quantity, 
determined by a ratio of coincidence counts, attenuation does not reduce it.

What we observe in Fig.4 could be considered a smooth transition 
between teleportation, when 
the teleporting OPO has strong squeezing, to continuous variable 
entanglement swapping, when the teleporting OPO has a conversion 
efficiency similar to that of the source OPO. The teleportation limit is 
characterized by the exact reproduction of the state of $B$ on $D'$. The 
beams $A$ and $D'$ are in the same polarization-number entangled state as 
$A$ and $B$ were originally. On the other hand, in the entanglement swapping 
limit, although every photon in $D'$ has become polarization entangled with a 
corresponding photon in beam $A$ the number entanglement has become 
strongly diluted. This is due to the effective attenuation 
which leaves many unpaired photons in 
beam $A$. The joint state of beams $A$ and $D'$ is now strongly mixed. This 
situation has been referred to by some authors as "a posteriori" 
teleportation \cite{branat}.

Homodyne detection losses will reduce and, for $\eta\le 1/S_{A,B}$, 
eventually destroy the
non-local effects. At the optimum gain condition (now 
$\lambda_{op}=\tanh\chi_{2}/\sqrt{\eta}$) this means the homodyne detection 
efficiencies must be better than about 83\%.
This limit is independent of the amount of 
squeezing. However, reducing 
the squeezing of OPO2 increases the effective attenuation at the 
optimum gain condition and hence reduces the coincidence 
count rate. As a result longer 
counting times are required to observe non-locality. This reduction in 
signal to noise is typical of entanglement swapping and is an 
unavoidable 
consequence of operating below unity gain \cite{ral98}. Never-the-less we 
believe an experimental demonstration is feasible with 
current technology. For example with $\eta=0.9$ and 50\% squeezing 
($\chi_{2}=0.34$) we find $S_{A,D'}=1.08$ with coincidence count rates reduced to 
about 10\% of their unteleported values.

In summary we have shown that it is possible to teleport the non-local 
correlations associated with number-polarization entanglement 
using a continuous variable scheme. The non-local correlations can be 
teleported for any level of squeezing in the teleporting OPO (OPO2). 
In general the best operating point for teleportation of the 
entanglement is where the output of the teleporter is simply an attenuated
version of the input beam. This operating point is clearly of importance 
for a large range of superposition and entangled state inputs.

We thank S.L.Braunstein for stimulating discussions. This work was 
supported by the Australian Research Council.

\begin{figure}
 \caption{Schematic of the teleportation arrangement.  OPO1 produces 
   pairs of photons with entangled polarizations in beams A and B. The
   polarization modes of beam B are separated, and teleported
   separately into beam D.
   The squeezed resource is provided by OPO2. 
   The polarization modes are swapped during teleportation, so
   a half wave plate is inserted to swap them back.  Finally,
   coincidence measurements are taken on modes A and D' to test for
   violation of the Clauser-Horne inequality.}
   \end{figure}

\begin{figure}
 \caption{Schematic of the interferometric modulation system (IMS).  
 The input beam is combined
   at a beamsplitter with a coherently related local oscillator. The 
   signal from the homodyne detectors is used to modulate
   the phases of each of the resulting beams, with a $\pi$ phase 
   shift between them.  The beams are then recombined. In the absence 
   of modulation the input emerges unchanged from the output port. 
   The phase modulation couples
   some of the intensity of the local oscillator into the output beam;
   in effect the signal is added to the amplitude of the beam.}
\end{figure}

\begin{figure}
 \caption{The variation of $S_{A,D'}$ with the
   polarizer angle ($\theta_{A}$) at unity gain. The other polarizers 
   are also varied such that the condition $\theta_{A}=-\theta_{D'}/2 
   =\theta_{A}'/3 $ is maintained whilst $\theta_{D'}'=0$. This 
   arrangement maximizes $S$.  The CH inequality is
   violated for $S>1$.  The two traces are for 99\% 
   and 80\% squeezing at OPO2. $\chi_{1}=0.1$}
\end{figure}

\begin{figure}
 \caption{The variation of $S_{A,D'}$ with the gain $\lambda$, 
 for squeezing of 10\%, 50\%, 80\% and 99\%.  Each graph has its maximum when 
 $\lambda = \tanh\chi_{2}$, in which case $S$ has the same value before and 
 after one of the beams is teleported. $\chi_{1}=0.1$}
\end{figure}

\end{document}